\documentclass[epj,twocolumn]{webofc}
\usepackage[varg]{txfonts}   

\providecommand{\openone}{\leavevmode\hbox{\small1\kern-3.8pt\normalsize1}}

\newcommand{\xt}{$(X\,T)$}
\newcommand{\tb}{$(T\,B)$}
\newcommand{\by}{$(B\,Y)$}

\wocname{EPJ Web of Conferences}
\woctitle{LHCP 2013}
\begin{document}
\title{Mixing with vector-like quarks: constraints and expectations}
%
%

\author{J. A. Aguilar-Saavedra\inst{1,2,3}\fnsep\thanks{\email{jaas@ugr.es}}}

\institute{Departamento de F\'{\i}sica Te\'orica y del Cosmos, 
Universidad de Granada, Granada, Spain
\and
           Departamento de Fisica, Universidade de Coimbra,  Coimbra, Portugal 
\and
           Instituto de F\'{\i}sica de Cantabria (CSIC-UC), Santander, Spain
          }

\abstract{%
We argue why vector-like quarks are usually expected to mix predominantly with the third generation, and discuss about the expected size of this mixing and its naturalness.
}
\maketitle
\section{Introduction}

New quarks beyond the three chiral families of the Standard Model (SM) have been searched at the Tevatron and now at the Large Hadron Collider (LHC). The most obvious\footnote{Here, obvious does not mean natural: a fourth generation of fermions includes a fourth neutrino heavier than half the mass of the $Z$ boson, that is, 11 orders of magnitude heavier than the other three neutrinos.} possibility of a fourth chiral generation is now excluded by indirect constraints and also by the agreement found for Higgs cross sections and decay branching ratios with the SM predictions. But this is not the only way to extend the quark sector. Indeed, vector-like quarks, namely multiplets whose left- and right-handed parts transform under the same $\text{SU}(2)_L$ representation, can exist and couple to the SM particles. In particular, they can be produced at the LHC and give a variety of clean characteristic signatures~\cite{AguilarSaavedra:2009es}.
It is usually assumed that vector-like quarks couple predominantly to the third SM generation. Here we discuss some motivations for this assumption, and investigate the expected size for that mixing.

\section{General mixing}
\label{sec:2}

If the scalar sector comprises only $\text{SU}(2)_L$ doublets, as it is suggested by the agreement of the measured Higgs boson properties with the SM predictions, there are only 
seven possibilities for vector-like multiplets coupling to the SM particles~\cite{delAguila:2000aa}
\begin{align}
& T_{L,R}^0 \,, \quad B_{L,R}^0 && \text{(singlets)} \,, \notag \\
& (X\,T^0)_{L,R} \,, \quad (T^0\,B^0)_{L,R} \,, \quad (B^0\,Y)_{L,R} && \text{(doublets)} \,, \notag \\
& (X\,T^0\,B^0)_{L,R} \,, \quad (T^0\,B^0\,Y)_{L,R}  && \text{(triplets)} \,,
\label{ec:mult.}
\end{align}
where a zero superscript is included in the weak eigenstates to distinguish them from the mass eigenstates. (This superscript will be omitted for brevity when it is clear from the context.) We will consider here minimal models with one additional multiplet. The resulting mass eigenstates are linear combinations of all weak eigenstates of the same charge, with mixing matrices that are determined from the diagonalisation of the mass matrix.  For example, for charge $2/3$ quarks, the addition of a new weak eigenstate $T^0$ results in a $4 \times 4$ mass matrix
\begin{equation}
\mathcal{L}_\text{mass} = - \left(\! \begin{array}{cc} \bar u_{Li}^0 & \bar T_L^0 \end{array} \!\right)
\left(\! \begin{array}{cc} y_{ij}^u \frac{v}{\sqrt 2} & y_{i4}^u \frac{v}{\sqrt 2} \\[1.5mm] y_{4j}^u \frac{v}{\sqrt 2} & M^0 \end{array} \!\right)
\left(\! \begin{array}{c} u^0_{Rj} \\[1.5mm] T^0_R \end{array}
\!\right)  +\text{H.c.} \,.
\label{ec:Lmass}
\end{equation}
We use a compact block notation where $u_{Li.Rj}^0$, $i,j=1,2,3$, are three-vectors of the SM fields $u_{L,R}^0,c_{L,R}^0,t_{L,R}^0$; $y_{ij}^u$ is the standard $3 \times 3$ matrix of Yukawa couplings and $y_{i4}^u$, $y_{4j}^u$ are matrices of Yukawa couplings with dimensions $3 \times 1$ and $1 \times 3$, respectively; $v=246$ GeV is the Higgs vacuum expectation value and $M^0$ is a bare mass term. When $T^0$ is a singlet or belongs to a triplet $y_{4j}^u=0$, while if it belongs to a doublet $y_{i4}^u=0$. The $4 \times 4$ mass matrix $\mathcal{M}^u$ appearing in Eq.~(\ref{ec:Lmass}) can be diagonalised by a biunitary transformation $U_L^u \mathcal{M}^u (U_R^u)^\dagger = \mathcal{M}^u_\text{diag}$, and the relation between mass and weak eigenstates is
$u_{L\alpha}^0 = (U_L^u)_{\alpha \beta} \, u_{L \beta}$, 
$u_{R\alpha}^0 = (U_R^u)_{\alpha \beta} \, u_{R \beta}$, with $\alpha,\beta=1-4$.
Exact expressions for $U_{L,R}^u$ are unmanageable, but simple ones can be obtained under the assumption that $M^0 \gg y \frac{v}{\sqrt 2}$, where $y$ denotes any of the Yukawa couplings. This assumption is well motivated by the lower limits on heavy quark masses set by LHC searches, which range between 600 and 800 GeV \cite{ATLAS:Ht,ATLAS:SS,ATLAS:Zb,ATLAS:WbWb}. Then, at first order, for singlets and triplets ($y_{4j}^u=0$) one has~\cite{Branco:1986my}
\begin{eqnarray}
U_L^u & = &
\left(\! \begin{array}{cc} \hat U_L^u & 0 \\[1.5mm] 0 & 1 \end{array} \!\right)
\left(\! \begin{array}{cc} \openone_{3 \times 3} & - y_{i4}^u \frac{v}{\sqrt 2M^0} \\[1.5mm] \left(y_{i4}^u \right)^\dagger \frac{v}{\sqrt 2M^0} & 1 \end{array} \!\right) \,, \notag \\
U_R^u & = &
\left(\! \begin{array}{cc} \hat U_R^u & 0 \\[1.5mm] 0 & 1 \end{array} \!\right) \,,
\label{ec:unit1}
\end{eqnarray}
where $\hat U_{L,R}^u$ are $3 \times 3$ unitary matrices that diagonalise the $3 \times 3$ submatrix of the lighter mass eigenstates. The mixing between the light and heavy eigenstates is given by the off-diagonal entries in the second factor of $U_L^u$, namely $-y_{i4}^u v/(\sqrt 2 M^0)$. The assumption
$M^0 \gg y \frac{v}{\sqrt 2}$ made for the approximate $4 \times 4$ diagonalisation implies that these mixings are small, which is in agreement with experimental constraints. For extra quark doublets ($y_{i4}^u=0$) one has instead
\begin{eqnarray}
U_L^u & = &
\left(\! \begin{array}{cc} \hat U_L^u & 0 \\[1.5mm] 0 & 1 \end{array} \!\right) \,,\notag \\
U_R^u & = &
\left(\! \begin{array}{cc} \hat U_R^u & 0 \\[1.5mm] 0 & 1 \end{array} \!\right)
\left(\! \begin{array}{cc} \openone_{3 \times 3} & - \left(y_{4j}^u \right)^\dagger \frac{v}{\sqrt 2M^0} \\[1.5mm] y_{4j}^u  \frac{v}{\sqrt 2M^0} & 1 \end{array} \!\right) \,.\notag \\
\label{ec:unit2}
\end{eqnarray}
The same formalism applies with trivial replacements to the down sector in the presence of an additional eigenstate $B^0$, with respective mixing matrices $U_{L,R}^d$, $\hat U_{L,R}^d$ and
$d_{L\alpha}^0 = (U_L^d)_{\alpha \beta} \, d_{L \beta}$,  
$d_{R\alpha}^0 = (U_R^d)_{\alpha \beta} \, d_{R \beta}$.

Let us now focus on the $3 \times 3$ mixing among SM quarks.
The standard $3 \times 3$ Cabibbo-Kobayashi-Maskawa matrix is $\hat U_L^u (\hat U_L^d)^\dagger$, and is experimentally measured to be nearly diagonal. This means that either $\hat U_L^u, \hat U_L^d \sim \openone_{3 \times 3}$, or 
$\hat U_L^u , \hat U_L^d \sim \hat U_L$, with $\hat U_L$ a unitary matrix with large off-diagonal entries. But the latter possibility is equivalent to the former with a redefinition of the initial weak eigenstates $(u_L^0\,d_L^0)$, $(c_L^0\,s_L^0)$, $(t_L^0\,b_L^0)$. So, in full generality one has $\hat U_L^u, \hat U_L^d \sim \openone_{3 \times 3}$. Furthermore, with a redefinition of the initial right-handed eigenstates one can also assume $\hat U_R^u = \hat U_R^d = \openone_{3\times 3}$. Hence, the strong hierarchy of SM quark masses $m_u \ll m_c \ll m_t$ in the up sector  implies $y_{11}^u \ll y_{22}^u \ll y_{33}^u$. Although we do not yet have a theory of flavour, it is natural to have the same hierarchy in the entries involving the new quarks
\begin{align}
& y_{14}^u \ll y_{24}^u \ll y_{34}^u  && \text{(singlets,triplets)}\,; \notag \\
& y_{41}^u \ll y_{42}^u \ll y_{43}^u  && \text{(doublets)}\,,
\end{align}
so that the mixing with the top quark, proportional to $y_{34}^u$ (or $y_{43}^u$ for doublets) is expected to be the largest one. The same argument, following from the mass hierarchy $m_d \ll m_s \ll m_b$, suggests that in the down sector the mixing with the bottom quark is dominant.

Aside from these plausibility arguments, there are stringent constraints on mixing from experimental data. The mixing of a vector-like quark with more than one light quark generates flavour-changing neutral interactions among the SM quarks~\cite{Branco:1986my,delAguila:1998tp}. Their non-observation then implies ---in our framework with only one additional multiplet--- that only one light quark can have significant mixing with the vector-like quark.
In the up sector the mixing with the top quark is less constrained than with the up and charm quarks~\cite{AguilarSaavedra:2002kr}, so also from an experimental point of view a larger mixing with the top quark is favoured. On the other hand, this does not happen in the down sector, where constraints on the bottom quark mixing (resulting from precise measurements of $Z \to b \bar b$ at LEP) are more stringent than the ones for the $d,s$ quarks. We note, however, that these constraints can be evaded in non-minimal models with more than one vector-like quark and, in particular, mixing with the light generations can be sizeable while fulfiling experimental constraints~\cite{Atre:2008iu}.

\section{Mixing with the third generation}

Assuming that the new states only mix with the third generation ones, the exact mass matrix diagonalisation is straightforward, and the non-trivial blocks of the $4 \times 4$ unitary matrices in Eqs.~(\ref{ec:unit1}), (\ref{ec:unit2}) read~\cite{Aguilar-Saavedra:2013qpa}
\begin{equation}
\left(\! \begin{array}{c} t_{L,R} \\ T_{L,R} \end{array} \!\right)
= \left(\! \begin{array}{cc} \cos \theta_{L,R}^u & -\sin \theta_{L,R}^u e^{i \phi_u} \\ \sin \theta_{L,R}^u e^{-i \phi_u} & \cos \theta_{L,R}^u \end{array}
\!\right)
\left(\! \begin{array}{c} t^0_{L,R} \\ T^0_{L,R} \end{array} \!\right)
\label{ec:mixu}
\end{equation}
for the up sector and
\begin{equation}
\left(\! \begin{array}{c} b_{L,R} \\ B_{L,R} \end{array} \!\right)
= \left(\! \begin{array}{cc} \cos \theta_{L,R}^d & -\sin \theta_{L,R}^d e^{i \phi_d} \\ \sin \theta_{L,R}^d e^{-i \phi_d} & \cos \theta_{L,R}^d \end{array}
\!\right)
\left(\! \begin{array}{c} b^0_{L,R} \\ B^0_{L,R} \end{array} \!\right)
\label{ec:mixd}
\end{equation}
for the down sector. In terms of the mass matrix elements, the mixing angles for singlets and triplets are~\cite{Aguilar-Saavedra:2013qpa}
\begin{eqnarray}
\tan 2 \theta_L^q & = & \frac{\sqrt{2} |y_{34}^q| v M^0}{(M^0)^2-(y_{33}^q)^2 v^2/2 - |y_{34}^q|^2 v^2/2}  \,, \notag \\
\tan \theta_R^q & = & \frac{m_q}{m_Q} \tan \theta_L^q \,,
\label{ec:angle1}
\end{eqnarray}
with  $(q,m_q,m_Q)=(u,m_t,m_T),(d,m_b,m_B)$ , and
\begin{eqnarray}
\tan 2 \theta_R^q & = & \ \frac{\sqrt{2} |y_{43}^q| v M^0}{(M^0)^2-(y_{33}^q)^2 v^2/2 - |y_{43}^q|^2 v^2/2} \,,  \notag \\
 \tan \theta_L^q & = & \frac{m_q}{m_Q} \tan \theta_R^q \,, 
\label{ec:angle2}
\end{eqnarray}
for doublets (see also~\cite{Dawson:2012di,Fajfer:2013wca}). Note that the diagonal entries of the mass matrices can be assumed real in full generality. The approximate expressions in Eqs.~(\ref{ec:unit1}), (\ref{ec:unit2}) can be recovered by expanding the trigonometric functions at first order in $v/M^0$. The masses of the third generation quark and its heavy partner are
\begin{eqnarray}
m_q & = & \cos \theta_L^q \cos \theta_R^q \, y_{33}^q \frac{v}{\sqrt 2} +\sin \theta_L^q \sin \theta_R^q \, M^0 \notag \\
& & - \cos \theta_L^q \sin \theta_R^q \, |y_{34}^q| \frac{v}{\sqrt 2} \,, \notag \\
m_Q & = & \sin \theta_L^q \sin \theta_R^q \, y_{33}^q \frac{v}{\sqrt 2} + \cos \theta_L^q \cos \theta_R^q \, M^0 \notag \\
& &  + \sin \theta_L^q \cos \theta_R^q \, |y_{34}^q| \frac{v}{\sqrt 2} \,
\label{ec:mass}
 \end{eqnarray}
for singlets and triplets. These equations can be inverted to write the moduli of the $2 \times 2$ mass matrix elements in terms of the two masses and the mixing angle,
\begin{eqnarray}
M^0 & = & \sqrt{m_Q^2 \cos^2 \theta_L^q + m_q^2 \sin^2 \theta_L^q} \,, \notag \\
|y_{34}^q| \frac{v}{\sqrt 2} & = & \frac{m_Q^2 - m_q^2}{M^0} \sin \theta_L^q \cos \theta_L^q \,, \notag \\
y_{33}^q \frac{v}{\sqrt 2} & = & \sqrt{m_Q^2 \sin^2 \theta_L^q + m_q^2 \cos^2 \theta_L^q - |y_{34}^q| \frac{v}{\sqrt 2}} \,. \notag \\
\label{ec:invert}
\end{eqnarray}
For doublets the expressions are analogous, and can be obtained from Eqs.~(\ref{ec:mass}), (\ref{ec:invert}) by interchanging $\theta_L^q$ and $\theta_R^q$ and replacing $y_{34}^q$ by $y_{43}^q$. We emphasise here that it is very convenient to give limits and parameterise observables such as cross sections in terms of the physical quark masses $m_q$, $m_Q$ and mixing angles (the angles in the left and right sectors are not independent, see Eqs.~(\ref{ec:angle1}), (\ref{ec:angle2})). Among other advantages, using this parameterisation one of the variables ---the third generation quark mass $m_q$--- is already fixed by experimental data. However, this parameterisation ``hides'' the fact that, for a heavy $m_Q$, a large mixing requires a large off-diagonal matrix element $y_{34}^q$ (or $y_{43}^q$), as it can be seen from Eqs.~(\ref{ec:angle1}), (\ref{ec:angle2}). We will discuss this issue for the up and down sectors in turn.

\subsection{Mixing in the up sector}

\begin{figure}[t]
\begin{center}
\begin{tabular}{c}
\hspace{-2.5mm} \includegraphics[width=7.2cm,clip=]{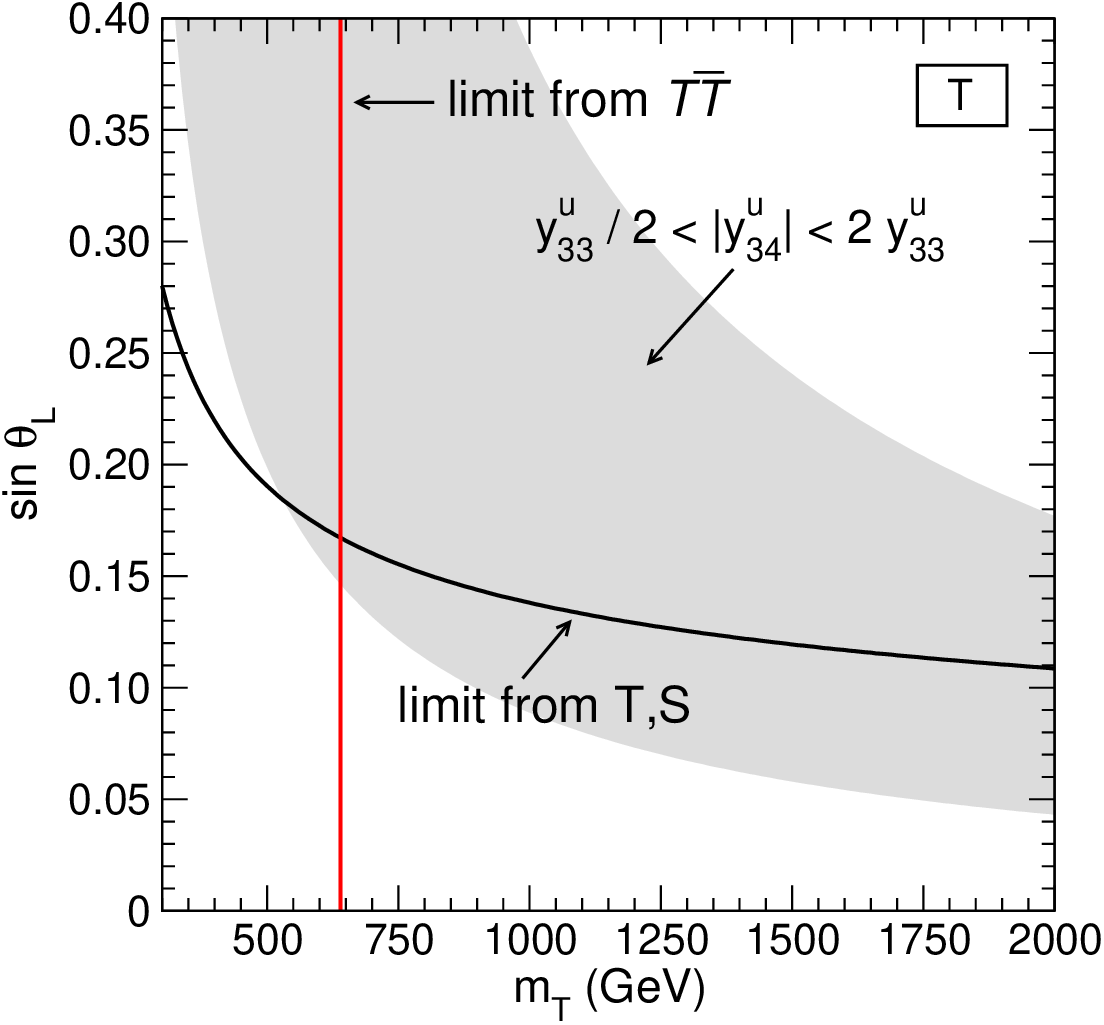} \\[2mm]
\includegraphics[width=7.2cm,clip=]{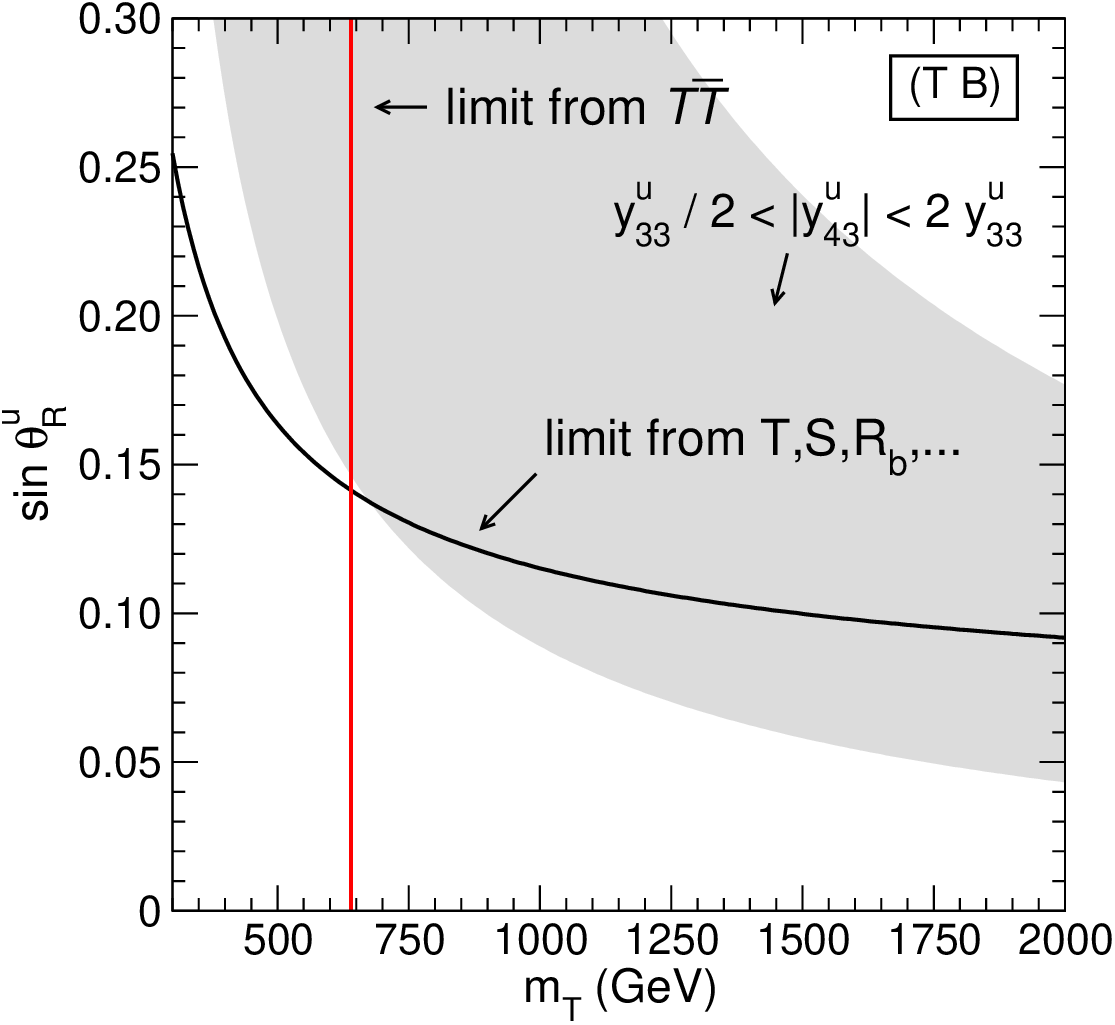} \\[2mm]
\includegraphics[width=7.2cm,clip=]{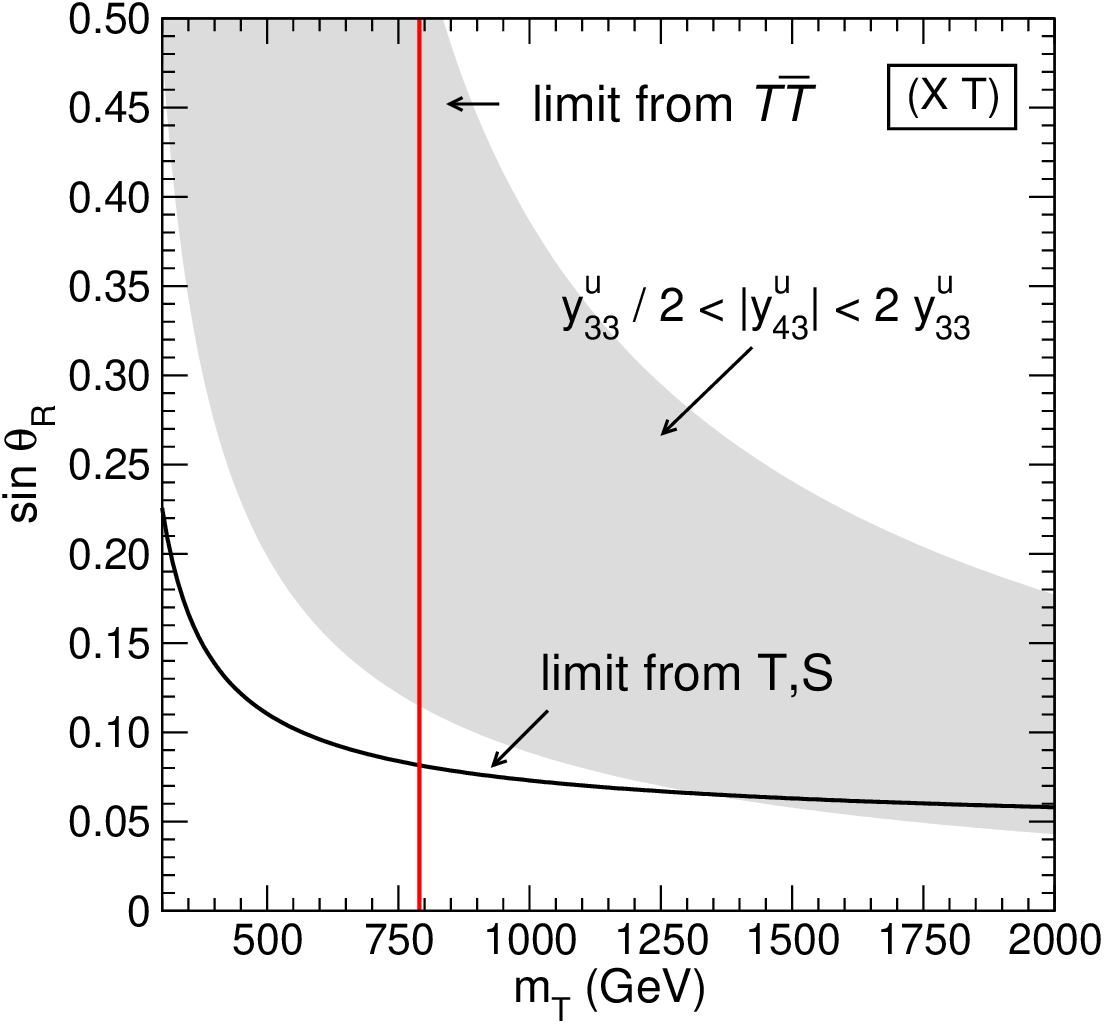}
\end{tabular}
\caption{Expected mixing of the top and its heavy partner for the $T$, \tb\ and \xt\ multiplets, for non-hierarchical Yukawa couplings. The upper limits on the mixing from precision measurements~\cite{Aguilar-Saavedra:2013qpa} and current mass limits from direct searches are also indicated.}
\label{fig:lim}
\end{center}
\end{figure}

It can be seen from Eqs.~(\ref{ec:mass}), that the top quark mass is $m_t \simeq y_{33}^u v/\sqrt 2$ to a good approximation (see below for a detailed derivation for the down sector). Then, the diagonal Yukawa coupling is large, $y_{33}^u \simeq 1$, just as in the SM. We can then ask ourselves which is the expected mixing in case that the diagonal and off-diagonal Yukawas are not hierarchical, taking for example $y_{34}^u$ (or $y_{43}^u$) between $y_{33}^u / 2$ and $2 y_{33}^u$. The result is given in Fig.~\ref{fig:lim}, where the upper limits on the mixing derived from the heavy quark contributions to oblique parameters $\mathrm{T},\mathrm{S}$ and $Z\to b \bar b$~\cite{Aguilar-Saavedra:2013qpa} are overlaid.

These plots clearly show that, unless the off-diagonal Yukawa coupling $y_{34}^u$ (or $y_{43}^u$) is significantly smaller than the top quark one, the mixing of the top quark with its heavy partner is expected to be close to its upper experimental limit. Moreover, in certain setups~\cite{Contino:2006nn} a larger off-diagonal coupling is expected. A detailed study of the phenomenological implications of a sizeable mixing for LHC searches and precision measurements of top couplings at the International Linear Collider (ILC) can be found in~\cite{Aguilar-Saavedra:2013qpa}.

\subsection{Mixing in the down sector}

In the down sector, non-hierarchical Yukawa couplings imply a very small mixing, due to the smallness of the bottom quark mass. However, this is not a necessary requirement and, in the same way as for the top quark, one can naturally have off-diagonal couplings of order unity~\cite{Chala:2013ega}. Then the question arises whether this does imply a fine tuning of the small $b$ quark mass, with cancellation of large contributions. We will show that this is not the case. Considering for definiteness the case of an extra $B$ doublet, and dropping the $d$ superscripts everywhere, one has from the first of Eqs.~(\ref{ec:mass})
\begin{eqnarray}
m_b & = & \cos \theta_L \cos \theta_R \, y_{33} \frac{v}{\sqrt 2} +\sin \theta_L \sin \theta_R \, M^0 \notag \\
&& - \sin \theta_L \cos \theta_R \, |y_{43}| \frac{v}{\sqrt 2} \,.
\label{ec:mb1}
 \end{eqnarray}
The mixing in the down sector is experimentally constrained to be small~\cite{Aguilar-Saavedra:2013qpa}. The limits are less restrictive for a \by\ doublet that explains the forward-backward asymmetry in $Z \to b \bar b$ with a mixing angle $\sin \theta_R \sim 0.15$. Therefore, one can use a second order Taylor expansion,
\begin{align}
& \sin \theta_R \simeq \theta_R \,,\quad \cos \theta_R \simeq 1 - \frac{1}{2} \theta_R^2 \,,\notag \\ 
& \sin \theta_L \simeq \frac{m_b}{m_B} \theta_R \,,\quad  \cos \theta_L \simeq 1 - \frac{1}{2} \frac{m_b^2}{m_B^2} \theta_R^2 \,, \notag \\ 
& M^0 \simeq m_B \left( 1-\frac{1}{2} \theta_R^2 \right) + \frac{m_b^2}{m_B^2} \theta_R^2 \,,\notag \\
& |y_{43}| \frac{v}{\sqrt 2} \simeq m_B \left( 1- \frac{m_b^2}{m_B^2} \right) \theta_L \,.
\end{align}
The second term in Eq.~(\ref{ec:mb1}) is the contribution to the $b$ quark mass from the bare mass term $M^0$, and equals $m_b \theta_R^2$, dropping higher orders in $\theta_R$. The third term is the contribution from the off-diagonal Yukawa coupling $y_{43}$, and equals $-m_b \theta_R^2(1-{m_b^2}/{m_B}^2)$. Both are quadratically suppressed, and their sum $m_b \theta_L^2 (m_b^2/m_B^2)$ is even more suppressed, so the $b$ quark mass is mainly given by the first term in Eq.~(\ref{ec:mb1}). Therefore, having a moderate mixing in the down sector can be natural provided one has a mechanism to yield an off-diagonal Yukawa coupling much larger than the bottom quark one.

\section*{Acknowledgements}
I thank M. P\'erez-Victoria, S. Heinemeyer and R. Benbrik for collaboration on this topic, and J. Santiago for discussions.
This work has been supported by MICINN by projects FPA2006-05294 and FPA2010-17915; by
Junta de Andaluc\'{\i}a (FQM 101 and FQM 6552) and by Funda\c c\~ao
para a Ci\^encia e Tecnologia~(FCT) project CERN/ FP/123619/2011.

\end{document}